\documentclass[12pt,letterpaper]{article}
\usepackage{graphicx}
\usepackage{amssymb}
\usepackage{amsmath}
\usepackage{enumerate}

\newcommand{\figscale}{0.7}
\newcommand{\del}{\partial}
\newcommand{\ppp}[2]{\frac{\partial #1}{\partial #2}}

\newcommand{\pppt}[3]{\frac{\partial^2 #1}{\partial #2 \partial#3 }}

\newcommand{\F}{{\mathcal{F}}}
\newcommand{\Tr}{{\textrm{Tr}}}
\newcommand{\image}{{\textrm{Im}}}
\newcommand{\W}{{\mathcal{W}}}
\newcommand{\bk}[1]{{\left\langle #1 \right\rangle}}

\renewcommand{\bar}[1]{{\overline{#1}}}

\newcommand{\sigmab}{{\overline{\sigma}}}
\newcommand{\delb}{{\overline{\del}}}
\newcommand{\thetab}{{\overline{\theta}}}
\newcommand{\Db}{{\overline{D}}}
\newcommand{\Fb}{{\overline{\F}}}
\newcommand{\alphad}{{\dot{\alpha}}}
\newcommand{\betad}{{\dot{\beta}}}

\begin{document}
%
%
\begin{titlepage}
\begin{flushright}
\normalsize
~~~~
OCU-PHYS 359\\
November, 2011\\
\end{flushright}

\vspace{15pt}

\begin{center}
{\LARGE  Proving the Absence of the Perturbative}\\~\\
{\LARGE Corrections to the $\mathcal{N}=2$ $U(1)$ K\"ahler Potential}\\~\\
{\LARGE Using the $\mathcal{N}=1$ Supergraph Techniques }\\
\end{center}

\vspace{23pt}

\begin{center}
{ Sho Deguchi$^{a}$\footnote{e-mail: deguchi@sci.osaka-cu.ac.jp}
}\\
%
\vspace{18pt}
%

$^a$ \it Department of Mathematics and Physics, Graduate School of Science\\
Osaka City University\\

\vspace{5pt}

3-3-138, Sugimoto, Sumiyoshi-ku, Osaka, 558-8585, Japan \\

\end{center}
%
\vspace{20pt}
\begin{center}
Abstract\\
\end{center}
Perturbative $\mathcal{N}=2$ non-renormalization theorem states that there is no perturbative correction to the K\"ahler potential $\int d^4\theta K\left(\Phi,\bar{\Phi}\right)$. We prove this statement by using the $\mathcal{N}=1$ supergraph techniques. We consider the $\mathcal{N}=2$ supersymmetric $U(1)$ gauge theory which possesses general prepotential $\F(\Psi)$.


\vfill

\setcounter{footnote}{0}
\renewcommand{\thefootnote}{\arabic{footnote}}

\end{titlepage}

\tableofcontents

\section{Introduction}
~~~In quantum field theory, there are, in general, many loop corrections to the effective action.
In the supersymmetric perturbative theory, there is non-renormalization theorem states that the corrections to the effective action take the form\cite{grs1979,pete,wsbg}
\begin{align}
\int d^{4\mathcal{N}}\theta \int d^4x_1\cdots d^4x_n 
	F_1(x_1,\theta,\thetab)\cdots F_n(x_n,\theta,\thetab)
        G(x_1,\cdots,x_n)~,
\end{align}
where the function $G$ is translationally invariant and $F$'s are product of superfields and their derivatives. There are no factors of $\square^{-1}(=[\del^m\del_m]^{-1})$ in the $F$'s, so the $d^{4\mathcal{N}}\theta$ integration cannot be converted into a $d^{2\mathcal{N}}\theta$ integration without generating additional spacetime derivatives.
In $\mathcal{N}=1$ supersymmetric theory, there is no loop correction to the effective superpotential $\int d^2\theta f(\Phi)$, while there is, in general, loop correction to the effective K\"ahler potential $\int d^4\theta K_{\textrm{eff}}(\Phi,\bar{\Phi})$.
On the other hand, in $\mathcal{N}=2$ supersymmetric $U(1)$ gauge theory, all of the loop corrections to the effective K\"ahler potential however cancel.
This is because of that the action of $\mathcal{N}=2$ supersymmetric theory takes the form of chiral ($\int d^2\theta_1d^2\theta_2\cdots$).
So there is no perturbative correction to the original terms of the action according to $\mathcal{N}=2$ non-renormalization theorem \cite{pete,wsbg}.
 The $\mathcal{N}=2$ non-renormalization theorem is given by using the unconstrained superfield \cite{pete,mezi1979} $V^{IJ}$($=\bar{V}^{IJ}$) as
\begin{align}
	\Psi\sim\Db^{IJ}\Db_{IJ} D_{KL} V^{KL}~,\label{eq:psi1}
\end{align}
where $I,J,K,L$ are global $SU(2)$ indices, and this superfield satisfies the chirality condition and reality condition as follows \cite{Grimm:1977xp}.
\begin{align}
D^I_\alpha\Psi=0~,~~~D^{IJ}\Psi=\bar{D}^{IJ}\bar{\Psi}.\label{eq:conditions}
\end{align}
 These conditions reduce the unwanted degree of freedom for the vector superfield.
On the other hand, it is not known how this theorem is proven by using the $\mathcal{N}=1$ formalism, where $\Psi$ is represented as
\begin{gather}
\Psi(\tilde{y},\theta,\tilde{\theta})
=\Phi(\tilde{y},\theta)
+i\sqrt{2}\tilde{\theta}^\alpha\W_\alpha(\tilde{y},\theta)
+\tilde{\theta}\tilde{\theta}\int d^2\thetab~\overline{\Phi}(\tilde{y}-i\tilde{\theta}\sigma\tilde{\thetab},\thetab)~.
\label{eq:n1form}
\end{gather}
where $\tilde{y}=x+i\theta\sigma\thetab+i\tilde{\theta}\sigma\bar{\tilde{\theta}}$, and $\tilde{\theta}$ is a second grassmann spinor. 
In this paper, we prove the cancellation of loop corrections to the effective K\"ahler potential.
In contrast to the simple model we considered in paper, there are 
many calculations of one- and two-loop effective K\"ahler potential in general $\mathcal{N}=1$ supersymmetric theories, which are 
performed in Ref. \cite{Grisaru:1996ve,Brignole:2000kg,Nibb}.

Outline of this paper is as follows.
In Sec.\ref{ss:model}, we represent $\mathcal{N}=2$ supersymmetric $U(1)$ gauge theory by using the $\mathcal{N}=1$ formalism. 
In Sec.\ref{ss:sr}, we construct super-Feynman rules and illustrate the correspondence between the supergraphs and the formulas by using an example.
Sec.\ref{ss:cs} is the central part in this paper, where we prove the cancellation upto the $n$-loop supergraphs.

\section{$\mathcal{N}=2$ supersymmetric $U(1)$ gauge theory in the $\mathcal{N}=1$ formalism}\label{ss:model}
We consider a theory, which can be represented by using $\mathcal{N}=2$ formalism as in Ref. \cite{Seiberg:1988ur}:
\begin{align}
S[\Psi,\bar\Psi]=\image\int d^4xd^2\theta d^2\tilde{\theta}~\F(\Psi)~,
\label{eq:lag2}
\end{align}
where $\F$ is a polynomial and we postulate that the imaginary part of the second derivative $\F$ is positive:
\begin{align}
\F(x)=\sum_{k=2}^{n} \frac{a_k x^k}{k!},~~~a_i\in \mathbb{C},~~ n\geqslant2~.
\end{align}
In the $\mathcal{N}=1$ SUSY formalism, the action is rewritten as:
\begin{align}
S[\Phi,\bar\Phi,V]=\image\int d^8z~\bar{\Phi}\ppp{\F(\Phi)}{\Phi}+\image\int d^6z\frac{1}{2!}\pppt{\F(\Phi)}{\Phi}{\Phi}\W^\alpha\W_\alpha.
\end{align}

To obtain the effective K\"ahler potential $\int d^4\theta~K_{\textrm{eff}}(\phi,\bar\phi)$ we decompose chiral superfield $\Phi$ into classical part $\phi$ and perturbative part $\Phi$, and perform the path integral respect to $\Phi$ and $V$ as follows.
\begin{gather}
\exp\left(i\int d^8z~K_{\textrm{eff}}(\phi,\bar\phi)\right)
=\int[d\Phi][d\bar\Phi][dV]\exp\left(iS[\phi+\Phi,\bar\phi+\bar\Phi,V]\right),\\
S\left[\phi+\Phi,\bar\phi+\bar\Phi,V\right]
=\image\int~d^8z\sum_{k=2}^{n}\frac{\F^{(k)}(\phi)}{k!}\left\{\bar\Phi\Phi^{k-1}+\frac{1}{2!}\Phi^{k-2}\left(D^\alpha V\right)\left(-\frac{1}{4}\Db^2D_\alpha V\right)\right\}~,
\end{gather} 
where we do not consider the derivatives of $\phi$ as in Ref.\cite{Nibb}, because we focus only on the effective K\"ahler potential.

\section{Super-Feynman Rules}\label{ss:sr}
In this section, we explain super-Feynman rules needed for deriving the effective action. We use the super-Feynman rules introduced by Grisaru, Ro\u{c}ek, and Siegel\cite{grs1979}.

\begin{enumerate}[I{)}]
\item For each line,
\begin{align}
\textrm{chiral line}:~~~
\raisebox{-.1ex}{\includegraphics[scale=\figscale]{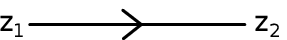}}
~~~=&~~~\frac{i}{g(\phi,\bar{\phi})\square}\delta^8(z_1-z_2)
~~~=~~~\langle\Phi(z_1)\bar{\Phi}(z_2)\rangle_{\textrm{GRS}},\\
\textrm{vector line}:~~~
\raisebox{-.0ex}{\includegraphics[scale=\figscale]{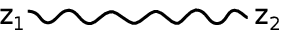}}
~~~=&~~~\frac{-i}{2g(\phi,\bar{\phi})\square}\delta^8(z_1-z_2)
~~~=~~~\langle V(z_1)V(z_2)\rangle,
\end{align}
where $g(\phi,\bar{\phi})\equiv\image(\F''(\phi))$ and $\square\equiv\del^m\del_m$.
\item For each $n$-point vertex ($n\geqslant 3$),
\begin{align}
\raisebox{-4.0ex}{\includegraphics[scale=0.6]{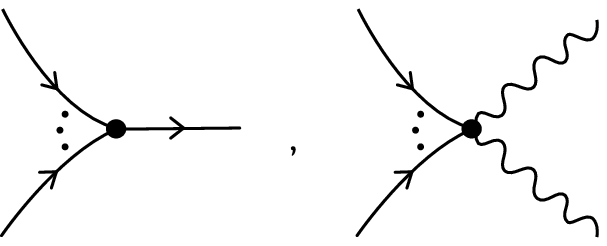}}
~~~~~=&~~~~~\int d^8z~\frac{1}{2}\F^{(n)}(\phi)~~,\\
\raisebox{-4.0ex}{\includegraphics[scale=0.6]{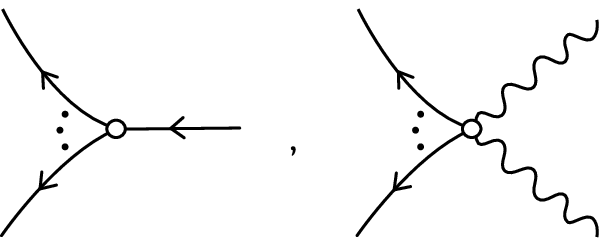}}
~~~~~=&~~~~~\int d^8z~(-\frac{1}{2}\bar{\F}^{(n)}(\bar{\phi}))~~,
\end{align}
where $\F^{(n)}(\phi)=\ppp{^n}{\phi^n}\F(\phi)$.
\item At each outgoing (or incoming) chiral line, we act $-\tfrac{1}{4}D^2$ (or $-\tfrac{1}{4}\Db^2$).
\item At each set of two vector lines connected with the holomorphic vertices, we act $-\tfrac{1}{4}\Db^2D^{\alpha}$ and $D_\alpha$. 
At each set of two vector lines connected with the anti-holomorphic vertices, we act $-\tfrac{1}{4}D^2\Db_{\dot{\alpha}}$ and $\Db^{\dot{\alpha}}$.\label{rule:vector}
\item $(-)$ sign is multiplied per one vector loop. This is because of anti-commutativity of operators acting at the vector lines in Rule \ref{rule:vector}.
\item Each supergraph is multiplied by symmetric factor.
\end{enumerate}

\subsection{Supergraph/Formula correspondence}
\begin{figure}[htb]
\centering
\includegraphics[scale=\figscale]{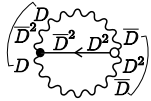}
\caption{one of the 2-loop supergraphs. (Lines connecting operators denote the pairs of the spinor contractions)}
\label{fig:example_graph}
\end{figure}

Here we consider a correspondence between the formulas and the supergraphs.
The following vacuum amplitude corresponds to the supergraph as in Fig.\ref{fig:example_graph}.
\begin{align}
&\left\langle
	\int dz_1^8~
		(\frac{1}{2}\F^{(3)}(\phi_1))
		\frac{1}{2!}
		\Phi_1
		(D_1^\alpha V_1)
		(-\frac{1}{4}\Db_1^2 D_{1\alpha} V_1)
	\right.\notag\\&~~~~~~~\cdot\left.
	\int dz_2^8~
		(-\frac{1}{2}\Fb^{(3)}(\bar\phi_2))
		\frac{1}{2!}
		\bar\Phi_2
		(\Db_{2\alphad} V_2)
		(-\frac{1}{4}D^2 \Db_{2}^{\alphad} V_2)
\right\rangle_{\textrm{connected}}\\
&=(-)\frac{1}{2!}\int dz_1^8\int dz_2^8(\frac{1}{2}\F^{(3)}(\phi_1))(-\frac{1}{2}\Fb^{(3)}(\bar\phi_2))
\notag\\&~~~~~~~\cdot
		\left((-\frac{1}{4}\Db_1^2)(-\frac{1}{4}D_2^2)
		\left\langle\Phi_1\bar\Phi_2\right\rangle_{\textrm{GRS}}\right)
		\left((-\frac{1}{4}\Db_1^2D_{1\alpha})(\Db_{2\alphad}) 
		\langle V_1V_2\rangle\right)
		\notag\\&~~~~~~~\cdot
		\left((-\frac{1}{4}D^2 \Db_{2}^{\alphad})(D_1^\alpha)
		\langle V_2V_1\rangle\right)
		\\
&=(-)\frac{1}{2!}\int dz_1^8\int dz_2^8(\frac{1}{2}\F^{(3)}(\phi_1))(-\frac{1}{2}\Fb^{(3)}(\bar\phi_2))
	\notag\\&~~~~~~~\cdot
		\left((-\frac{1}{4}D^2)(-\frac{1}{4}\Db^2)
		\frac{i}{g\square}\delta^8(z_1-z_2)\right)
		\left((-\frac{1}{4}\Db_1^2 D_{1\alpha})(\Db_{2\alphad}) 
		\frac{-i}{2g\square}\delta^8(z_1-z_2)\right)
		\notag\\&~~~~~~~\cdot
		\left((-\frac{1}{4}D^2 \Db_{2}^{\alphad})(D_1^\alpha)
		\frac{-i}{2g\square}\delta^8(z_1-z_2)\right),
		\label{eq:2loop2}
\end{align}
where the over all minus sign comes from the existence of one vector loop. The front coefficients $\frac{1}{2!}$ are symmetric factors come from exchanging of the vector superfield lines.

\section{Cancellation of supergraphs}\label{ss:cs}
In this section, we prove the cancellation of loop corrections to the effective K\"ahler potential $K(\Phi,\bar{\Phi})$. 
Cancellation of the lowest order correction is proven in Sec.\ref{ss:oneloop}.
Cancellation of the \emph{holomorphic} loops is proven in Sec.\ref{ss:hloop}.
For simplicity, the vertices are redefined as the summation of the contracted holomorphic trees as described in Sec.\ref{ss:cht},
and it is given how operators act typically in chiral line and vector line except common operators in Sec.\ref{ss:dcv}.
To begin with cancellation of 2-loops is proven in Sec.\ref{ss:2loop}.
Final Sec.\ref{ss:nloop} is the essence of our paper where we prove that all of the $n$-loop corrections to the effective K\"ahler potential cancel.

\subsection{1-loop}\label{ss:oneloop}
Quadratic term of the action include the superfields called: chiral superfields $\Phi$,$\bar{\Phi}$, vector superfield $V$, Faddev-Pappov ghosts $C$,$C'$,$\bar{C}$,$\bar{C'}$,  and Nielsen-Kallosh ghost $\chi$,$\bar{\chi}$ as follow.
$$
S_{\textrm{quad}}=\int d^8z \left(g\Phi\bar{\Phi}+gV\square V - C'\bar{C}-\bar{C'}C - 2g\bar{\chi}\chi\right),
$$
where $g=\image\F^{(2)}(\phi)$. Lowest-order correction takes the form:
$$
z_{\textrm{quad}} =\int[d\Phi][d\bar{\Phi}][dV][dC][d\bar{C}][dC'][d\bar{C'}][d\chi][d\bar{\chi}]\exp(iS_{\textrm{quad}}).
$$
In this case, vector term do not depend on $g$ as follow.
$$
\int[dV]\exp\left(\int d^8z~[gV\square V+\varepsilon V\square V]\right)
=\int[dV]\exp\left(\int d^8z~[gV\square V]\right).
$$
This is because of that $\varepsilon$-vertex do not have operators like $D$ and $\bar{D}$. So perturbative calculation derive the delta function $\delta^2(\theta-\theta)\delta^2(\bar\theta-\bar\theta)=0$. 
Chiral term cancel with term of Nielsen-Kallosh ghost which is fermionic, and Faddev-Poppov ghost term do not depends on $g(\phi,\bar{\phi})$.
Lowest-order correction is absent.

\subsection{Holomorphic loop}\label{ss:hloop}
Here we mean by holomorphic loop a loop constructed only with holomorphic vertices.
The holomorphic loops cancel each other by the following summation.
\begin{figure}[htb]
\centering
\includegraphics[scale=\figscale]{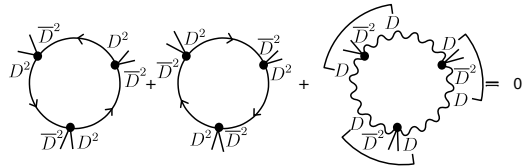}
\caption{holomorphic loop cancelation}
\label{fig:loop_sum}
\end{figure}

We confirm the Fig.\ref{fig:loop_sum} 
by partial integrations respect to $D$ and $\bar{D}$ in the following procedure.

Summation of first two supergraphs in Fig.\ref{fig:loop_sum} gives the following amplitude, here we omit the common factors($\bar{D}^2$ and $\sim\square^{-1}$) between the chiral loops and the vector loop.
$$
2\cdot(-\frac{1}{4}D^2~\delta^8)
	   (-\frac{1}{4}D^2~\delta^8)
	   (-\frac{1}{4}D^2~\delta^8)
$$
Third supergraph gives
\begin{align*}
&(-)\cdot(-)^3(-\frac{1}{2})^3\cdot
	(D_\alpha D^\beta~\delta^8)
	(D_\beta D^\gamma~\delta^8)
	(D_\gamma D^\alpha~\delta^8)\\
&~~~~~=
(-)\cdot(-)^3(-\frac{1}{2})^3\cdot
	(-\frac{1}{2}\delta_\alpha^\beta D^2~\delta^8)
	(-\frac{1}{2}\delta_\beta^\gamma D^2~\delta^8)
	(-\frac{1}{2}\delta_\gamma^\alpha D^2~\delta^8)\\
&~~~~~=
(-)2\cdot
	(-\frac{1}{4}D^2~\delta^8)
	(-\frac{1}{4}D^2~\delta^8)
	(-\frac{1}{4}D^2~\delta^8),
\end{align*}
where $(-)^3$ comes from change of variables $z$ to $z'$, $(-\frac{1}{2})^3$ is ratio between propagators of chiral superfields and vector superfields, and we have used $D_\alpha D^\beta=-\frac{1}{2}\delta_\alpha^\beta D^2$.

\subsection{Contractions of the holomorphic trees}\label{ss:cht}
\begin{figure}[htb]
\centering
\includegraphics[scale=\figscale]{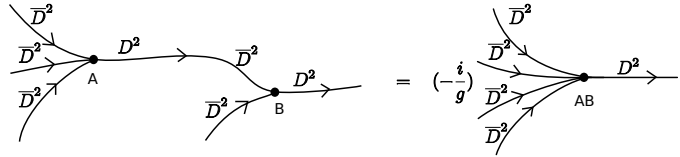}
\caption{Chiral tree contraction}
\label{fig:tree_press_c}
\end{figure}

Any chiral line connecting two holomorphic vertices is eliminated by integrating out the delta function $\delta^8(z-z')$ on the lines as in Fig.\ref{fig:tree_press_c}. 
Typically the delta functions at these chiral lines take the form:
$$
(-\frac{1}{4}D^2)(-\frac{1}{4}\Db^2)(\frac{i}{g\square})\delta^8.
$$
By partial integral respect to $-\frac{1}{4}D^2$ from the another line, we can act the additional operator to this chiral line as follow.
\begin{align}
(-\frac{1}{4}D^2)(-\frac{1}{4}\Db^2)(-\frac{1}{4}D^2)(\frac{i}{g\square})\delta^8
=-(-\frac{1}{4}D^2)(\frac{i}{g})\delta^8.
\end{align}
Finally $-\frac{1}{4}D^2$ is back to the initial line by partial integral.
So we can integrate out the delta functions of inner lines.
 
\begin{figure}[htb]
\centering
\includegraphics[scale=\figscale]{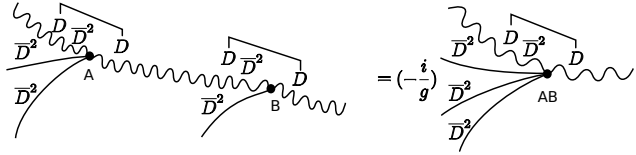}
\caption{vector tree contraction}
\label{fig:tree_press_v}
\end{figure}

On the other hand, any vector line connecting two holomorphic vertices is eliminated as in Fig.\ref{fig:tree_press_v} in the same way.
Typically the delta functions at these vector lines take the form:
$$
(-)(D_\alpha)(D^\beta(-\frac{1}{4}\Db^2))(\frac{-i}{2g\square})\delta^8.
$$
By partial integral, $-\frac{1}{4}D^2$ acts as follows.
\begin{align}
&\rightarrow
(-)(-\frac{1}{4}\Db^2)(D_\alpha)(D^\beta(-\frac{1}{4}\Db^2))(\frac{-i}{2g\square})\delta^8
\\
&=
(-)(-\frac{1}{4}\Db^2)(-\frac{1}{2}\delta_\alpha^\beta D^2)(-\frac{1}{4}\Db^2)(\frac{-i}{2g\square})\delta^8
\\
&
=
(-\frac{1}{4}\Db^2)(-\frac{1}{4}D^2)(-\frac{1}{4}\Db^2)(\frac{i}{g\square})\delta^8
\\
&=
(-\frac{1}{4}\Db^2)(-\square)(\frac{i}{g\square})\delta^8
\\
&\rightarrow
-\frac{i}{g}\delta^8.
\end{align}

\begin{figure}[htb]
\centering
\includegraphics[scale=\figscale]{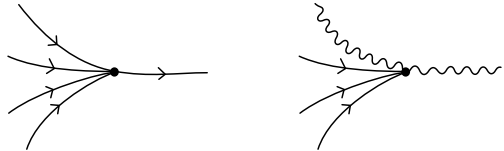}
\caption{These vertices are made from the contracted holomorphic trees}
\label{fig:tree_press}
\end{figure}

In this way we obtain vertices made from the contracted holomorphic trees as in Fig.\ref{fig:tree_press}, and we use these vertices as fundamental ones.
Then the additional super-Feynman rule arise, that the holomorphic and anti-holomorphic vertices can not connect with the holomorphic and anti-holomorphic ones, respectively. This rule prevents the overlap of summation of trees. 

\subsection{Differences between chiral lines and vector lines}\label{ss:dcv}

As our purpose in this paper is seeing the cancellation between the loop corrections, we focus on the differences between chiral lines and vector lines. 
There are two types of $n$-point vertices that include two vector line and that include only chiral lines. But these coefficients are the same.
However, ways of acting momentum operator $P_m=i\del_m$ are different.
Difference between chiral lines and vector lines are following.

\begin{figure}[htb]
\centering
   \includegraphics[scale=\figscale]{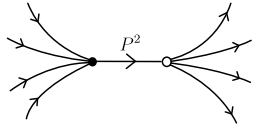}
   \caption{typical way of how operators act at vertices including only chiral lines.}
\label{fig:bold_tree}
\end{figure}

Each vertex including only chiral lines have one opposite directional chiral line. After little calculation like in Sec.\ref{ss:hloop}, we can see that $P^2$ act at these opposite directional chiral lines as in Fig.\ref{fig:bold_tree}, where we omitted operators which are common to chiral lines and vector lines. 

\begin{figure}[htb]
\centering
   \includegraphics[scale=\figscale]{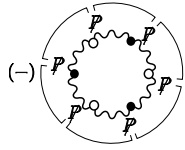}
  \caption{typical way of how operators act at vertices including vector lines. (\emph{vector line is always closed. See Sec.\ref{ss:sr}})}
\label{fig:tvloop}
\end{figure}

Each vertex including vector lines is connected with two vector lines. In the same way, we can see that ${\not}P$ act at these vector lines as in Fig.\ref{fig:tvloop}, where minus sign comes from existence of one vector line loop (See Sec.\ref{ss:sr}) and we omitted operators which are common to chiral lines and vector lines.
(Slashed vector operators, for example, ${\not}A$ denote $A_m\sigma^m$ or $A_m\sigmab^m$. ${\not}A{\not}B{\not}C\cdots$ denote the product of alternate $\sigma^m$'s and $\sigmab^m$'s as $(A_mB_nC_k\cdots)(\sigma^m\sigmab^n\sigma^k\cdots)$.)

\subsection{2-loop}\label{ss:2loop}
\begin{figure}[htb]
\centering
\includegraphics[scale=\figscale]{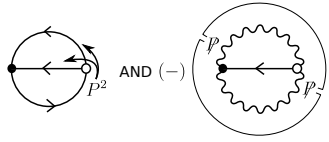}
\caption{2-loop graphs}
\label{fig:loop_two_cv}
\end{figure}

Holomorphic loop diagrams cancel as discussed in Sec.\ref{ss:hloop}.
Similarly, non-holomorphic loop diagrams cancel.
In this subsection, we consider the two loop diagrams as in Fig.\ref{fig:loop_two_cv}.
There are other diagrams including tadpole.
However we do not consider these diagrams in this subsection.
In Sec.\ref{ss:nloop}, we show the cancellation, including the tadpoles. 
The cancellation of these two diagrams is shown by partial integral respect to $P_m$ ---(Here, we used $P^2=\frac{1}{2} \Tr({\not}P{\not}P)=\frac{1}{2} P_mP_n\Tr(\sigma^m\sigmab^n)$).

\subsection{$n$-loop}\label{ss:nloop}
\begin{figure}[htb]
\centering
\includegraphics[scale=\figscale]{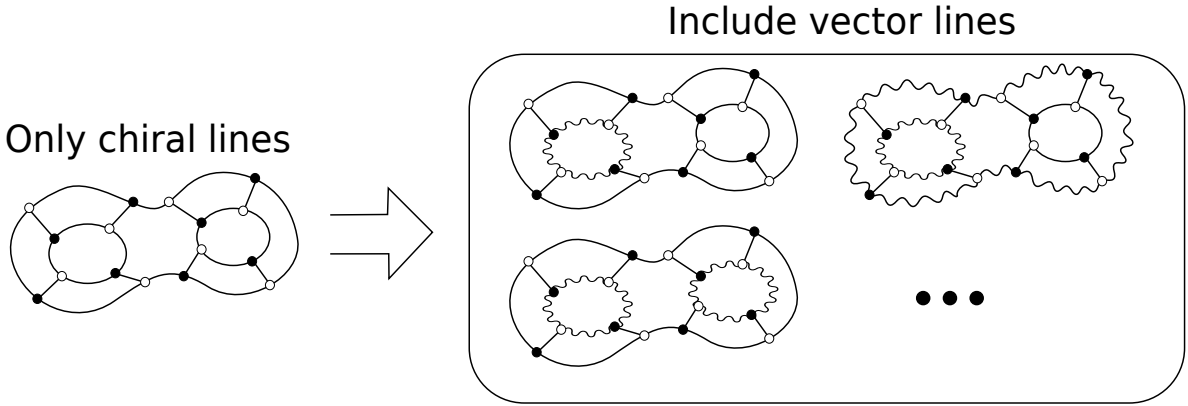}
\caption{transformation of supergraph from one including only chiral lines to other including vector lines.}
\label{fig:cancellation}
\end{figure}

Difference between chiral lines and vector lines is summarized in Sec.\ref{ss:dcv}.
Then we show that difference is mostly only the way of how operator ${\not}P$ are acting.
In this way, we now focus on a supergraph including only chiral lines as left hand side in Fig.\ref{fig:cancellation}.
This supergraph can be transformed into supergraphs including vector like lines as right hand side in Fig.\ref{fig:cancellation} by appropriate partial integral respect to ${\not}P$. 
Then we notice that these supergraphs cancel for opposite sign.

\begin{figure}[htb]
\centering
\includegraphics[scale=\figscale]{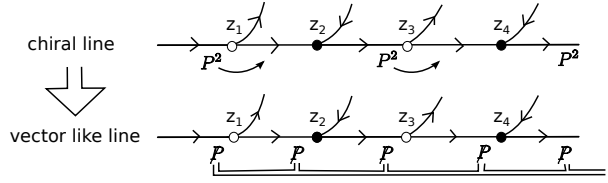}
\caption{transform chiral line to vector like line by partial integral respect to ${\not}P$.}
\label{fig:chain}
\end{figure}

The transformations from chiral lines to vector like lines are given by appropriate partial integral respect to ${\not}P$ as follow.
\begin{align}
P_1^2P_3^2\cdots=\tfrac{1}{2}\Tr({\not}P^2_1{\not}P^2_3\cdots)\longrightarrow\tfrac{1}{2}\Tr({\not}P_1{\not}P_2{\not}P_3{\not}P_4\cdots),\label{eq:maketrace}
\end{align}
where subscript $1,2,\cdots$ denote position $z_{1},z_2,\cdots$, and we used $P^21_{2\times2}={\not}P^2$.
We can describe above state by using supergraphs as in Fig.\ref{fig:chain}.
Here we are focusing on only one route of partial integral for simplicity, so we in fact need to consider about another routes by choosing another lines for acting ${\not}P$.

\begin{figure}[htb]
\centering
\includegraphics[scale=\figscale]{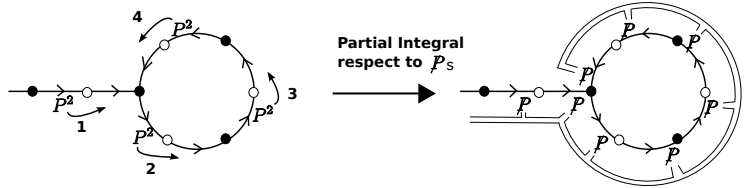}
\caption{dead ends type 1}
\label{fig:dead1}
\end{figure}

\begin{figure}[htb]
\centering
\includegraphics[scale=\figscale]{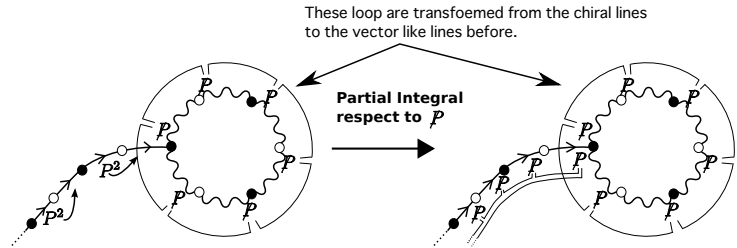}
\caption{dead ends type 2}
\label{fig:dead2}
\end{figure}

However this procedure will comes to two types of dead ends that there is no next $P^2$ to do the partial integration as in Fig.\ref{fig:dead1} and Fig.\ref{fig:dead2}, respectively.

\begin{figure}[htb]
\centering
\includegraphics[scale=\figscale]{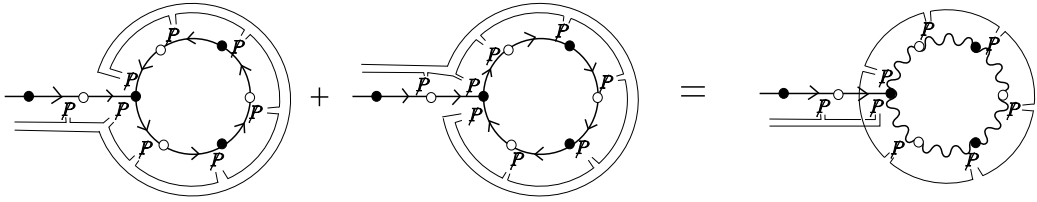}
\caption{split the trace}
\label{fig:loop_cut}
\end{figure}

In the dead ends type 1 as in Fig.\ref{fig:dead1}, we can split the trace by considering opposite directional chiral loop as in Fig.\ref{fig:loop_cut},
where we use the identity:
\begin{align}
({\not}A_1{\not}A_2\cdots{\not}A_{2n})
+
({\not}A_{2n}\cdots{\not}A_2{\not}A_{1})
~=~
2\Tr({\not}A_1{\not}A_2\cdots{\not}A_{2n})1_{2\times2}.
\label{eqn:trace_cut}
\end{align}
And so we can continue the procedure by turning ${\not}P$ into another line.

\begin{figure}[htb]
\centering
\includegraphics[scale=\figscale]{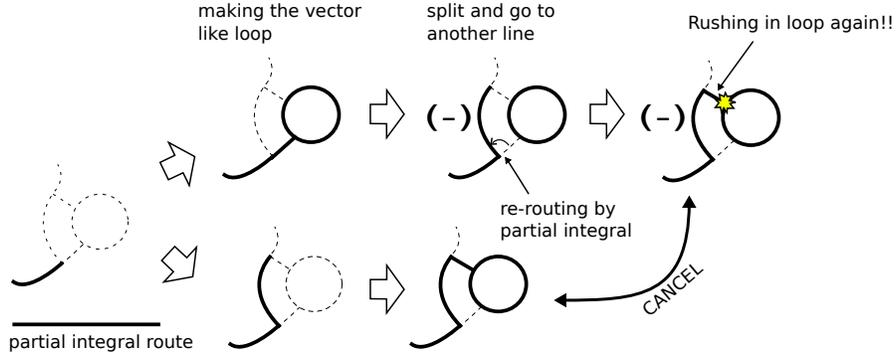}
\caption{hitting routes cancel.}
\label{fig:rush}
\end{figure}

In the dead ends type 2 as in Fig.\ref{fig:dead2}, the partial integral procedure hits into the vector like loop previously transformed by appropriate partial integrations.
But these partial integral routes cancel with the other routes as in Fig.\ref{fig:rush}.
Thus we can neglect these hitting routes, and continue the partial integral procedure.

\begin{figure}[htb]
\centering
\includegraphics[scale=\figscale]{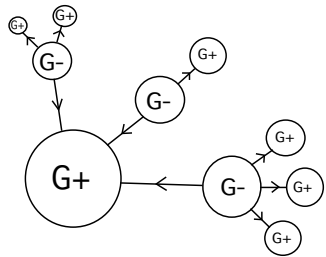}
\caption{$G+$ is a region of the supergraphs that we can reach along the lines with the direction of chiral line. On the other hand, $G-$ is a region of the supergraphs that we can reach along the lines with the opposite direction to the direction of chiral line.}
\label{fig:reach}
\end{figure}

Ultimately partial integral procedure back to start point.
Then many supergraphs including vector like loops are generated. 
However there is a situation that we can not generate all of the supergraphs including vector line by doing the single above procedure. This is because of that, in the single procedure, we consider the partial integral along the one side direction of chiral line or this opposite direction. So there are regions that we can not reach by the single partial integral as in Fig.\ref{fig:reach}.
But we can cover all supergraphs including vector lines by executing the above procedure repeatedly from the different regions and by flipping the direction of partial integral procedure for the direction of chiral lines.
  
Then all of the supergraphs including vector superfield are generated with opposite sign, so all of the supergraphs cancel, including tadpoles.

\section{Conclusion and Outlook}
We have proven
that all of the loop corrections to the effective K\"ahler potential, in
$\mathcal{N}=2$ $U(1)$ gauge theory, cancel each other
by using $\mathcal{N}=1$ supergraph techniques instead of
$\mathcal{N}=2$ ones.

As for applications, 
the extension to the $U(1)^N$ case is rather trivially carried out
 while  the extension to the $U(N)$ case requires more efforts.
On the other hand, $\mathcal{N}=2$ supersymmetry is broken in nature and
there is $U(N)$ gauge theory in which  $\mathcal{N}=2$
supersymmetry is partially and spontaneously broken \cite{apt1995,FIS1,FIS2,FIS3}.
The action of this theory possesses the superpotential term albeit being 
$\mathcal{N} =2$ supersymmetric and in the presence of such superpotential term
 the proof in this paper no longer holds.

\section*{Acknowledgments}
The author acknowledges helpful discussions with Hiroshi Itoyama as well as continual encouragement and careful reading of the manuscript. He also thanks Yukinori Yasui for helpful comments. He is supported by the scholarship of Graduate School of Science in Osaka City University for doctoral students.

\appendix
\section{Conventions and Notations}
We use metric $\eta_{mn}=diag(+,-,-,-)$, and spacetime coordinate $x^m$, where $m,n,\cdots=0,1,2,3$.
We use also grassmann valued spinor $\theta_\alpha$ and this complex conjugate $\thetab_\alphad$, these indices are raised by $\epsilon^{\alpha\beta}$ ($\epsilon^{12}=\epsilon_{21}=1,~\epsilon^{21}=\epsilon_{12}=-1$, and other components are zero).
Pauli matrix takes the form:
\begin{gather}
\sigma^1=
\begin{pmatrix}
0&1\\
1&0
\end{pmatrix}
~,~~~
\sigma^2=
\begin{pmatrix}
0&-i\\
i&0
\end{pmatrix}
~,~~~
\sigma^3=
\begin{pmatrix}
1&0\\
0&-1
\end{pmatrix},
\\
\sigma^m=(I,\vec{\sigma})~~,~~~~~
\sigmab^m=(I,-\vec{\sigma})~~,~~~~~
{\not}A:=A_m\sigma^m~\textrm{or}
~A_m\sigmab^m~~.
\end{gather}
Raising and lowering indices, and contraction:
\begin{align}
\psi^\alpha=\epsilon^{\alpha\beta}\psi_\beta~~,~~~
\overline{\psi}^\alphad=\epsilon^{\alphad\betad}\overline{\psi}_{\betad}~~,~~~
\psi\lambda=\psi^\alpha\lambda_\alpha~~,~~~
\bar{\psi}\bar{\lambda}=\bar{\psi}_\alphad\bar{\lambda}^\alphad~~.
\end{align}
The derivatives respect to grassmann number: 
\begin{gather}
\del_\alpha=\ppp{}{\theta^\alpha}~,~~~
\del^\alpha=\ppp{}{\theta_\alpha}~,~~~
\overline{\del}^{\dot{\alpha}}=\ppp{}{\overline{\theta}_{\dot{\alpha}}}~,~~~
\overline{\del}_{\dot{\alpha}}=\ppp{}{\overline{\theta}^{\dot{\alpha}}}~,~~~\\
\del_\alpha\theta^\beta=\delta_\alpha^\beta~,~~~
\del^\alpha\theta_\beta=\delta^\alpha_\beta~,~~~
\overline{\del}^{\dot{\alpha}}\overline{\theta}_{\dot{\beta}}=\delta^{\dot{\alpha}}_{\dot{\beta}}~,~~~
\overline{\del}_{\dot{\alpha}}\overline{\theta}^{\dot{\beta}}=\delta_{\dot{\alpha}}^{\dot{\beta}}~.
\end{gather}
The covariant derivatives for N=1 rigid superspace are given by:
\begin{align}
D_\alpha&=\del_\alpha+i\sigma^m_{\alpha\dot{\beta}}\overline{\theta}^{\dot{\beta}}\del_m~~,\\
D^\alpha&=-\del^\alpha-i\overline{\theta}_{\dot{\beta}}\sigmab^{m\dot{\beta}\alpha}\del_m~~,\\
\Db^\alphad&=\delb^\alphad+i\sigmab^m{}^{\alphad\beta}\theta_\beta\del_m~~,\\
\Db_\alphad&=-\delb_\alphad-i\theta^\beta\sigma^m_{\beta\alphad}\del_m~~.
\end{align}
Their anti-commutation relations are:
\begin{align}
\{D_\alpha,D_\beta\}=0=\{\Db^\alphad,\Db^\betad\}~~,~~~
\{D_\alpha,\Db_\betad\}=-i2\sigma^m_{\alpha\betad}\del_m
\end{align}
Translation operator $P_m=i\del_m$.
Projection operator:
\begin{gather}
P_0=\frac{D^\alpha\overline{DD}D_\alpha}{8\square}
=\frac{\Db^\alphad DD\Db_\alphad}{8\square}
~,~~~P_{+}=-\frac{\overline{DD}DD}{16\square}
~,~~~P_{-}=-\frac{DD\overline{DD}}{16\square}\notag\\
P_0+P_{+}+P_{-}=1\label{eq:projection}
\end{gather}
$P_0$ is sum of two projection operators $\frac{D^1\overline{DD}D_1}{8\square}$ and $\frac{D^2\overline{DD}D_2}{8\square}$.
Property of Chiral superfield:
\begin{gather}
\Db_\alphad\Phi=D_\alpha\overline{\Phi}=0~~,~~~
P_0\Phi=P_0\overline{\Phi}=0~~,~~~
P_+\Phi=\Phi~,~~~P_-\overline{\Phi}=\overline{\Phi}~~.
\end{gather}
Integral respect to Grassmann number
\begin{gather}
\int d^2\theta \theta\theta=1
~,~~~
\int d^2\thetab \thetab\thetab=1\\
\int d^6z=\int d^4xd^2\theta
~,~~~
\int d^6\overline{z}=\int d^4x d^2\thetab
~,~~~
\int d^8z=\int d^4xd^4\theta=\int d^4xd^2\theta d^2\thetab
\end{gather}
\subsection{$\mathcal{N}=2$ formalism}
Most naive extension of $\mathcal{N}=1$ to $\mathcal{N}=2$ take the form:
\begin{align}
\{D^I,\bar{D}_J\}&=-i\sigma^m\del_m\delta^I_J~~,\\
\{D^I,D_J\}&=0~~.
\end{align} 
where $I,J,K,L$ are global $SU(2)$ indices. 
\begin{gather}
\bar{D^I}=\bar{D}_I~,\\
D_I=\epsilon_{IJ}D^J~,~~~D^I=\epsilon^{IJ}D_{J}~,\\
\bar{D}_I=\epsilon_{IJ}\bar{D}^J~,~~~\bar{D}^I=\epsilon^{IJ}\bar{D}_{J}~.
\end{gather}
A little useful notation:
$$
	D^{IJ}:=D^ID^J~,~~~\bar{D}^{IJ}:=\bar{D}^I\bar{D}^J
$$
$$
D^+:=D^{11},~
D^0:=\sqrt{2}D^{12},~
D^-:=D^{22}.
$$
$$
D^+D_+=D^0D_0=D^-D_-=D_{11}D_{22}=\frac{1}{4}D^{IJ}D_{IJ}.
$$
Projection operators:
$$
Q^{i}{}_{j}:=\frac{D^i\bar{D}_{+}\bar{D}_{+}D_j}{(-16\square)^2}~,~~~~
\bar{Q}^{i}{}_{j}:=\frac{\bar{D}^i{D}_{+}{D}_{+}\bar{D}_j}{(-16\square)^2}
~,~~~~(i,j=+,-,0).
$$
Property of $Q^{i}{}_{j}$:
\begin{align}
Q^{i}{}_{j}Q^{k}{}_{l}&=\delta^{k}_{j}Q^{i}{}_{l},\\
\sum_i Q^i{}_i&\neq 1,\\
Q^{i}{}_{j}&=\bar{Q}^{i}{}_{j}.\label{eq:reality}
\end{align}
this is because of:
\begin{align}
Q^i{}_j\Leftarrow \frac{D^i\bar{D}^+\bar{D}_+D^+D_+\bar{D}_j}{(-16\square)^3}\Rightarrow\bar{Q}^i{}_j.
\end{align}
Chirality condition and reality condition:
$$
\bar{D}^I\Psi=0~,~~~
D^{i}\Psi=\bar{D}^{j}\bar{\Psi}~.
$$
Then $\Psi$ is written as follow.
\begin{align}
\Psi
\sim\bar{D}^+\bar{D}_+D_iV^i,\label{eq:psiN2}
\end{align}
where $V^+=V^{11},~V^{0}=\sqrt{2}V^{12},~V^-=V^{22}$, and $V^i$ is real which mean $V^i=\bar{V}^i$.
Chirality condition is trivial and so confirm reality condition.
$$
D^{i}\bar{D}^+\bar{D}_+D_jV^{j}=\bar{D}^{i}D^+D_+\bar{D}_jV^j~\leftrightarrow~eqn.\eqref{eq:reality}
$$
$\Psi$ in $\mathcal{N}=1$ formalism is
$$
\Phi(\tilde{y},\theta)+i\sqrt{2}\tilde{\theta}\mathcal{W}(\tilde{y},\theta)+\tilde{\theta}\tilde{\theta}G(\tilde{y},\theta),
$$
where $\theta:=\theta^1$, $\tilde{\theta}:=\theta^2$, and $\tilde{y}:=x+i\theta\sigma\bar{\theta}+i\tilde{\theta}\sigma\bar{\tilde{\theta}}~(\bar{D}^I\tilde{y}=0)$.
Form of $G$ is decided by reality condition.
First term is written by using unconstraint superfield as follows.
\begin{align}
\Phi(\tilde{y},\theta)
&=\bar{D}\bar{D}f(x+i\tilde{\theta}\sigma\bar{\tilde{\theta}},\theta,\bar{\theta})\\
&=\frac{1}{16}\bar{D}\bar{D}\bar{\tilde{D}}\bar{\tilde{D}}\tilde{D}\tilde{D}\tilde{\theta}\tilde{\theta}\bar{\tilde{\theta}}\bar{\tilde{\theta}}f(x,\theta,\bar{\theta}).\label{eq:first}
\end{align}
For second term:
\begin{align}
i\sqrt{2}\tilde{\theta}\mathcal{W}(\tilde{y},\theta)
&=-\frac{1}{32}\tilde{\theta}^\alpha\bar{D}\bar{D}D_\alpha\bar{\tilde{D}}\bar{\tilde{D}}\tilde{D}\tilde{D}\tilde{\theta}\tilde{\theta}\bar{\tilde{\theta}}\bar{\tilde{\theta}}g(x,\theta,\bar{\theta})\\
&=\frac{1}{16}\bar{D}\bar{D}\bar{\tilde{D}}\bar{\tilde{D}}\tilde{D}D\tilde{\theta}\tilde{\theta}\bar{\tilde{\theta}}\bar{\tilde{\theta}}g(x,\theta,\bar{\theta}),\label{eq:second}
\end{align}
where $g$ is imaginary superfield ($g=-\bar{g}$).
For third term:
\begin{align}
\tilde{\theta}\tilde{\theta}G(\tilde{y},\theta)
&=-\frac{1}{4}\tilde{\theta}\tilde{\theta}\bar{D}\bar{D}h(x+i\tilde{\theta}\sigma\bar{\tilde{\theta}},\theta,\bar{\theta})\\
&=-\frac{1}{64}\tilde{\theta}\tilde{\theta}\bar{D}\bar{D}\bar{\tilde{D}}\bar{\tilde{D}}\tilde{D}\tilde{D}\tilde{\theta}\tilde{\theta}\bar{\tilde{\theta}}\bar{\tilde{\theta}}h(x,\theta,\bar{\theta})\\
&=\frac{1}{16}\bar{D}\bar{D}\bar{\tilde{D}}\bar{\tilde{D}}\tilde{\theta}\tilde{\theta}\bar{\tilde{\theta}}\bar{\tilde{\theta}}h(x,\theta,\bar{\theta})\label{eq:third}
\end{align}
Superfield $\Psi$ (Eqn.\eqref{eq:psiN2}) is 
\begin{align}
\Psi\sim \bar{D}\bar{D}\bar{\tilde{D}}\bar{\tilde{D}}(\tilde{D}\tilde{D}V^{11}+2D\tilde{D}V^{12}+DDV^{22})
\end{align}
Because of the superfield $V^{IJ}=\bar{V}^{IJ}$,
$$
V^{11}=\bar{V}^{11}=\bar{V_{11}}=\bar{V^{22}}~,~~~
V^{12}=\bar{V}^{12}=\bar{V_{12}}=-\bar{V^{12}}~,~~(V^{12}=V^{21}).
$$
Then $\Psi$ is written by using the arbitrary superfields $F,G(=-\bar{G})$:
\begin{align}
\Psi\sim \bar{D}\bar{D}\bar{\tilde{D}}\bar{\tilde{D}}(\tilde{D}\tilde{D}F+2D\tilde{D}G+DD\bar{F})~,~~~G=-\bar{G}
\end{align}
Compare this formula with sum of Eqn.\eqref{eq:first}\eqref{eq:second}\eqref{eq:third}:
$$
\Psi\sim
\bar{D}\bar{D}\bar{\tilde{D}}\bar{\tilde{D}}(\tilde{D}\tilde{D}\tilde{\theta}\tilde{\theta}\bar{\tilde{\theta}}\bar{\tilde{\theta}}f+2D\tilde{D}\tilde{\theta}\tilde{\theta}\bar{\tilde{\theta}}\bar{\tilde{\theta}}g+\tilde{\theta}\tilde{\theta}\bar{\tilde{\theta}}\bar{\tilde{\theta}}h)~,~~~(g=-\bar{g})
$$
So third term is decided as follow.
\begin{align}
h&=DD\bar{f}=\bar{\Phi}~,\\
\tilde{\theta}\tilde{\theta}G(\tilde{y},\theta)
&=-\frac{1}{4}\tilde{\theta}\tilde{\theta}\bar{D}\bar{D}\bar{\Phi}(\bar{y},\bar{\theta})\\
&\sim\tilde{\theta}\tilde{\theta}\int d^2\bar{\theta}~\bar{\Phi}(\bar{y},\bar{\theta})\\
\bar{y}&=\bar{x+i\theta\sigma\bar{\theta}}=x-i\theta\sigma\bar{\theta}
\end{align}

\section{Gauge Fixing}
\subsection*{Gauge invariance}
Here we define gauge transformation:
\begin{align}
	\Phi&\rightarrow \Phi'=\Phi\notag\\
	V&\rightarrow V'=V+i\Lambda-i\overline{\Lambda}
	\label{eq:gauge_trans}\\
	\W&\rightarrow \W'=\W\notag
\end{align}
where $\Lambda$ is Chiral superfield.
\subsection*{Derivation of Gauge Fixing term}
We considered a model of the following action.
\begin{align}
S=\image\int d^8z~\bar{\Phi}\F'(\Phi)+\image\int d^6z\frac{1}{2}\F''(\Phi)\W^\alpha\W_\alpha.
\end{align}
the action is invariant under $U(1)$ gauge transformation of eqn.\eqref{eq:gauge_trans}:
\begin{align}
S[\Phi,V]=S[\Phi,V']~.
\end{align}
Gauge fixing with Gauge condition $G$:
\begin{align}
Z&=\int[d\Lambda][d\overline{\Lambda}]|\delta(G(\Lambda,\overline{\Lambda}))|^2\det\left[\ppp{(G,\overline{G})}{(\Lambda,\overline{\Lambda})}\right]
\int[dV][d\Phi][d\overline{\Phi}]\exp\left(iS[\Phi,V']\right)\\
&=\int[d\Lambda][d\overline{\Lambda}]|\delta(G(\Lambda,\overline{\Lambda}))|^2\Delta_{FP}
\int[dV][d\Phi][d\overline{\Phi}]\exp\left(iS[\Phi,V']\right)\label{eq:gauge_fixed_S},
\end{align}
where $V'=V+i\Lambda-i\Lambda$, and the determinant at first line is Faddev-Pappov determinant $\Delta_{FP}$
Faddev-Pappov determinant $\Delta_{FP}$ represented by Faddev-Pappov ghost as follow:
\begin{align}
\Delta_{FP}
&=\int [dC'][d\overline{C'}][dC][d\overline{C}]~\exp
\left[
 -i\int d^6z~
 \left(
		C'
		\left.\ppp{G}{\Lambda}\right|_{G=0}
		\!\!\!\!\!\!\!\!C
		+
		C'
		\left.
			\ppp{G}{\overline{\Lambda}}
		\right|_{G=0}
		\!\!\!\!\!\!\!\!\overline{C}~~
 \right)
\right.
\notag\\&~~~~~~~
\left.
 -i\int d^6\overline{z}~
 \left(
		\overline{C'}
		\left.
		  \ppp{\overline{G}}{\Lambda}
		\right|_{\overline{G}=0}
		\!\!\!\!\!\!\!\!C
		+
		\overline{C'}
		\left.
			\ppp{\overline{G}}{\overline{\Lambda}}
		\right|_{\overline{G}=0}
		\!\!\!\!\!\!\!\!\overline{C}~~
 \right)
\right]
\end{align}
Assume gauge condition $G$ as:
\begin{align}
G=F-\frac{1}{4}\Db\Db V'
\end{align}
then
\begin{gather}
\Delta_{FP}
=\int [dC'][d\overline{C'}][dC][d\overline{C}]~\exp
\left[
 -i\int d^8z~
 \left(
		C'
		\overline{C}
		+
		\overline{C'}
		C
 \right)
\right]\\
S_{FP}=-\int d^8z~
 \left(
		C'
		\overline{C}
		+
		\overline{C'}
		C
 \right).
\end{gather}
Chiral superfield $F$ set the position of gauge fixing.
We integral with respect to $F$, and introduce Nielsen-Kallosh(NK) ghost $\chi$ for normalization.
\begin{gather}
\int [dF][d\overline{F}][d\chi][d\overline{\chi}]~\exp\left(-i\int d^8z~\left(2g\overline{F}F+2g\overline{\chi}\chi\right)\right)\sim 1\label{eq:FFXX}\\
S_{GF}=-\int d^8z~2g\overline{F}F
~,~~~
S_{NK}=-\int d^8z~2g\overline{\chi}\chi
\end{gather}
We obtain gauge fixing term by eliminating $F$:
\begin{align}
	S_{GF}&=-i\int d^8z~\frac{1}{8}gDDV\Db\Db V
\intertext{by partial integral:}
	&=-i\int d^8z~\frac{1}{16}gV(\Db\Db DD+DD\Db\Db) V\\
	&=i\int d^8z~gV(P_++P_-)\del^2 V\label{eq:GF}
\end{align}
Total action is
\begin{gather}
S_K+S_G+S_{GF}+S_{NK}+S_{FP}~.
\end{gather}
\section{Propagator of Chiral superfield}
\begin{align}
S&=\int d^8z ~\bar{\Phi}\Phi\\
&\rightarrow \bar{\Phi}\Phi
\end{align}
Introducing source $J,\bar{J}$:
\begin{align}
	S&=\bar{\Phi}\Phi+J\Phi+\bar{J}\bar{\Phi}\\
	&=\frac{1}{2}(\Phi~~\bar{\Phi})
	\begin{pmatrix}
		0&1\\1&0
	\end{pmatrix}
	\begin{pmatrix}
	\Phi\\\bar{\Phi}
	\end{pmatrix}
	+
	(\Phi~~\bar{\Phi})
	\begin{pmatrix}
	\frac{DD}{4\del^2}&0\\
	0&\frac{\bar{DD}}{4\del^2}
	\end{pmatrix}
	\begin{pmatrix}
	J\\
	\bar{J}
	\end{pmatrix}\\
	&=\frac{1}{2}(\Phi~~\bar{\Phi})
	A
	\begin{pmatrix}
	\Phi\\\bar{\Phi}
	\end{pmatrix}
	+
	(\Phi~~\bar{\Phi})B
	\begin{pmatrix}
	J\\
	\bar{J}
	\end{pmatrix}\\
	&=
	\bar{\Phi}'\Phi'
	-
	\frac{1}{2}
	(J~~\bar{J})
	BA^{-1}B
	\begin{pmatrix}J\\\bar{J}\end{pmatrix}\\
	&=
	\bar{\Phi}'\Phi'
	+
	\frac{1}{2}
	(J~~\bar{J})
	\begin{pmatrix}
	0&\frac{1}{\del^2}\\\frac{1}{\del^2}&0
	\end{pmatrix}
	\begin{pmatrix}J\\\bar{J}\end{pmatrix}\\
	&=
	\bar{\Phi}'\Phi'
	+
	\bar{J}\frac{1}{\del^2}J
\end{align}
Grisaru, Ro\v{c}ek and Siegel(GRS) propagator is:
\begin{align}
\boxed{
	\bk{\bar{\Phi}\Phi}_{\textrm{GRS}}=\frac{i}{\del^2}~,~~
	\bk{\Phi\Phi}_{\textrm{GRS}}=
	\bk{\bar{\Phi}\bar{\Phi}}_{\textrm{GRS}}=0~.
}
\end{align}
\section{Propagator of Vector superfield}
\begin{align}
	S&=\int d^6z~\frac{1}{2}\W\W\\
	&\rightarrow 
		\frac{1}{2}
		\left(-\frac{1}{4}\bar{DD}D^\alpha V\right)
		\left(-\frac{1}{4}\bar{DD}D_\alpha V\right)\\
	&=V\frac{D\bar{DD}D}{8} V
\intertext{Add gauge fixing term:}
	&\rightarrow V\del^2V.
\end{align}
Introducing source $j$:
\begin{align}
S&=V\del^2V+jV\\
&=V'\del^2V'-\frac{1}{4}j\frac{1}{\del^2}j.
\end{align}
Hence GRS vector propagator is:
\begin{align}
\boxed{
\bk{VV}_{\textrm{GRS}}=-\frac{1}{2}\frac{i}{\del^2}~.
}
\end{align}
Note that there is coefficient difference between Chiral propagator and vector one, with ratio $1:-\frac{1}{2}$.


\end{document}